\documentclass[aps,prb,twocolumn,superscriptaddress,amsmath,amssymb,revsymb,showpacs]{revtex4}

\usepackage{graphicx}
\usepackage{dcolumn}
\usepackage{bm}
\usepackage{color}

\newcommand{\be}{\begin{eqnarray}}
\newcommand{\ee}{\end{eqnarray}}
\newcommand{\nn}{\nonumber}
\newcommand{\lb}{\label}
\newcommand{\bo}{\boldsymbol}
\begin{document}
\title{Superconductivity in repulsively interacting 
fermions on a diamond chain:\\ 
flat-band induced pairing}

\author{Keita Kobayashi}%
\affiliation{%
CCSE, Japan Atomic Energy Agency, Kashiwa, Chiba
 277-0871, Japan}%
\author{Masahiko Okumura}%
\affiliation{%
CCSE, Japan Atomic Energy Agency, Kashiwa, Chiba
 277-0871, Japan}%
\author{Susumu Yamada}%
\affiliation{%
CCSE, Japan Atomic Energy Agency, Kashiwa, Chiba
 277-0871, Japan}%
\affiliation{%
Computational Materials Science Research Team, RIKEN AICS, Kobe, Hyogo
650-0047, Japan 
}%
\author{Masahiko Machida}%
\affiliation{%
CCSE, Japan Atomic Energy Agency, Kashiwa, Chiba
 277-0871, Japan}%

\author{Hideo Aoki}%
\affiliation{%
Department of Physics, University of Tokyo, Hongo, Tokyo 113-0033, Japan}
\affiliation{Electronics and Photonics Research Institute, 
Advanced Industrial Science and Technology (AIST), 
Tsukuba, Ibaraki 305-8568, Japan}%
\date{\today}

\begin{abstract}
To explore whether a flat-band system can accommodate 
superconductivity, we consider repulsively interacting fermions on 
the diamond chain, a simplest quasi-one-dimensional 
system that contains a flat band. 
Exact diagonalization and the density-matrix renormalization group (DMRG) 
are used to show that we have a significant 
binding energy of a Cooper pair with a 
long-tailed pair-pair correlation in real space when the total 
band filling is slightly below $1/3$, where a filled 
dispersive band interacts with the flat band that is 
empty but close to $E_F$. 
Pairs selectively formed across the outer sites 
of the diamond chain are responsible for the pairing 
correlation.  
At exactly $1/3$-filling 
an insulating phase emerges, where 
the entanglement spectrum indicates the particles 
on the outer sites are highly entangled and topological. 
These come from a peculiarity of the 
flat band in which ``Wannier orbits" are not 
orthogonalizable.  
\end{abstract}

\maketitle

{\it Introduction---} 
While fascinations with unconventional 
superconductivity arising from electron correlation continue to 
increase, as 
exemplified by the high-$T_C$ cuprates and iron-based superconductors, 
a next question to ask is whether 
there exists an avenue where we have superconductivity with 
another pairing mechanism.  Namely, in the  
superconductivity in correlated electron systems, 
the standard viewpoint is that the interaction mediated by 
spin fluctuations glues the electrons into anisotropic pairs 
such as d-wave or s$_{+-}$, where 
the nesting of the Fermi surface dominates the 
fluctuation, hence the superconductivity.  
To look for a different class of models, one 
intriguing direction is to consider correlated systems 
on flat-band lattices 
that contain dispersionless band(s) in their 
band structure.  This is because, 
regardless of the Fermi energy residing on or 
off the flat band, we cannot define the Fermi 
surface for the flat band.   In other words, we cannot 
apply, in one-dimensional cases, Tomonaga-Luttinger picture 
for the states around $E_F$  
even with multichannel g-ology unlike the case of ladders.  
Thus, if 
superconductivity does arise, 
this might harbor a mechanism in which the flat band plays 
a role distinct from the conventional, nesting-dominated 
boson-exchange mechanisms.


In the field of ferromagnetism, on the other hand, 
there is a long history 
for the study of flat-band ferromagnetism \cite{Lieb,Mielke,Tasaki}\,,   
which is distinct from the conventional (Stoner) 
ferromagnetism.  
The ferromagnetic ground state is rigorously 
shown for arbitrary repulsive interaction $0<U\leq \infty$ 
when the flat band is half-filled.  
The flat-band lattice models are 
conceived as Lieb model\cite{Lieb} with different numbers of 
A and B sublattice sites, or Mielke and Tasaki models\cite{Mielke,Tasaki} such as 
kagome lattice.  A speciality of these flat-band lattices 
appears as an anomalous situation that 
Wannier orbitals cannot be 
orthogonalized, which is called the connectivity condition 
for the density matrix\cite{tasakiPTP}. 
This immediately dictates that 
the flat band arises from interferences, hence 
totally different from the atomic 
(zero-hopping) limit, and indeed the flat-band models are 
necessarily multi-band systems, where the flat band(s) 
coexist with dispersive ones.  
Flat-band systems are not merely a theoretical curiosity, 
but candidate systems have been considered\cite{yamada}. 
Also, recent developments in cold-atom Fermi gases on 
optical lattices are a promising arena, where 
Lieb\cite{TakahashiLieb} and kagome~\cite{Kagome} lattices 
are already discussed. 

Thus the flat-band system provides a unique playground, 
because the correlation effects 
should be strong for the flat bands 
(as briefly described in Supplementary Material C), 
but also because of 
the above-mentioned unusual structure of the density 
matrix (or strongly interfering wave functions).  
We can thus envisage dramatic, possibly non-perturbative 
phenomena from the electron-electron interaction on these 
macroscopically degenerate manifolds of single-particle states. 
Beside the ferromagnetism, the flat band systems have attracted recent attentions for possible realization of topological insulators with non-trivial Chern numbers~\cite{Tang,Sun,Neupert,Li,Tovmasyan}.  
The next goal, in our view, is to realize superconductivity 
in flat-band systems.  We shall show here that there are indeed 
signatures for pairing for repulsively interaction 
electrons in a one-dimensional flat-band lattice.

Theoretically, exploration of superconducting phases 
in flat-band systems is quite challenging, since 
correlation effects become 
even more difficult to fathom for the 
flat bands than in ordinary ones\cite{RVB}. 
Thus far, possibility of pair formation on flat bands has 
been examined by several authors. 
Pairing of two fermions on diamond chain with $\pi$-flux inserted 
was discussed by Vidal et al\cite{Vidal}.  
Kuroki et al have considered a cross-linked ladder that contains 
wide and narrow (or flat) bands in the 
context of the high-$T_C$ cuprate ladder compound,\cite{Kuroki} 
and have shown that superconducting $T_C$ estimated 
from the fluctuation exchange approximation (FLEX) 
is much higher than in usual lattices when 
$E_F$ is just above the flat band that 
interspears the dispersive one.   There, 
virtual pair scatterings between the dispersive 
and fully-filled flat bands are suggested to cause the high $T_C$. 
Pair formation has also been discussed for 
bose Hubbard model on cross-linked ladders,\cite{Takayoshi,Huber} 
where 
a large pair hopping gives rise to the 
emergence of a superfluid phase (pair Tomonaga-Luttinger liquid) 
overlapping with a Wigner-solid region 
in the phase diagram. Namely, in flat-band 
systems, not only pair hopping amplitudes can be large, but 
also diagonal orders tend to coexist (rather than compete) 
with superfluids. 
These results suggest that the flat bands 
may be indeed a good place to look for pair condensates.

\begin{figure}[h]
\begin{center}
\includegraphics[width=1.0\linewidth]{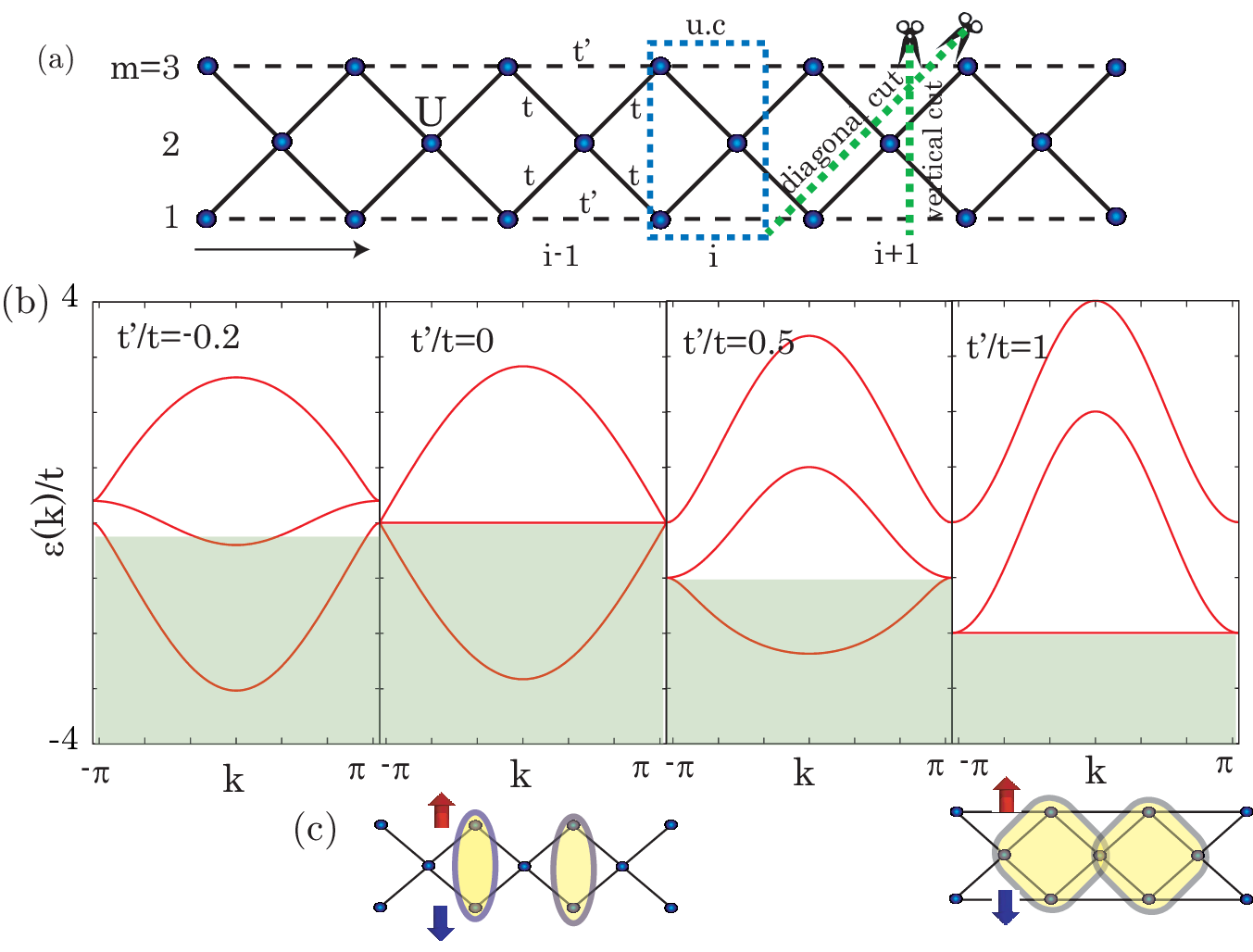}
\end{center}
\caption{(Color Online) 
(a) Hubbard model on a diamond chain with $t$ ($t'$) 
the nearest-neighbor (inter-apex) hoppings, 
$m$ labeling the leg while $i$ the unit cell. 
Also shown are two types of cuts 
(vertical and diagonal), which are used 
in DMRG calculation of the entanglement entropy. 
(b) Band structures in the 
noninteracting case $(U=0)$ for various values of $t'$, 
with shaded areas indicating the 1/3 filling. 
(c) Orbits considered here for the flat band at $t'/t=0$ or $1$. 
}
\label{Fig1}
\end{figure}

This has motivated us here to explore 
superconductivity for a repulsive fermionic Hubbard 
model on flat band systems.  As a model we take 
a simplest 
possible, quasi-1D lattice comprising a chain of 
diamonds as depicted in Fig.\ref{Fig1}(a).  
We shall show that for $E_F$ close to but slightly 
below the flat band 
(with the filling of the whole bands slightly 
below 1/3), attractive binding energies appear. 
Concomitantly, the pair-pair correlation becomes 
long-tailed in real space  at these 
band fillings.

{\it Model and methods ---} As methods for calculation we opt for exact 
diagonalization and the density-matrix renormalization group 
(DMRG) 
that can deal with strong correlation, since 
the correlation phenomena on flat bands may well 
call for such non-perturbative methods.
For the position of the Fermi energy, $E_F$, we focus on 
the regime where the flat band is empty.   
This choice comes from the following observation.  
When the flat band is half filled, the ground state 
is ferromagnetic.  When $E_F$ is 
shifted but still on the flat band, the diverging 
density of one-electron states is expected to 
give rise to large self-energy corrections, which 
should be detrimental to superconductivity.  
When the flat band is empty with $E_F$ 
residing in a dispersive band, this problem 
can be resolved, with virtual processes between 
the dispersive and flat bands still at work.  
For bipartite lattices such as the diamond chain, 
the empty flat band is equivalent 
to fully-filled flat band due to 
an electron-hole symmetry 

It is desirable to have, on top of $E_F$, another 
control parameter about the flat band.  So here 
we introduce a 
hopping $t'$ between the adjacent apex sites of diamonds (Fig.\ref{Fig1}(a)).  
For $t'=0$ the lattice (a Lieb model) 
is bipartite with the 
flat band as a middle one 
in this three-band system.  As we increase $t'$ the 
bands are deformed, until 
in the limit $t'/t=1$ the bottom band becomes 
flat (a Mielke model).  Thus we can examine how the pairing behaves 
as we change $t'=0 \rightarrow 1$. 
We then calculate  the binding energy of pairs 
with the exact diagonalization (ED), and 
pair-pair (and other) correlation functions with 
the DMRG~\cite{DMRG1,DMRG2,Okumura2009,Okumura2011}.

We take the conventional 
Hubbard Hamiltonian on the diamond chain (Fig.\ref{Fig1}(a)), 
\be
&&H=H_{\rm kin}+H_{\rm int}\,, \lb{eq:hami}\\
&&H_{\rm kin}=t\sum_{i,\sigma=\uparrow\downarrow}
c_{2,i,\sigma}^\dagger
\sum_{m=1,3} (c_{m,i,\sigma}+c_{m,i+1,\sigma}) \nn \\
&&\qquad\quad+ t'\sum_{i,\sigma=\uparrow\downarrow}\sum_{m=1,3}
c_{m,i,\sigma}^{\dagger}c_{m,i+1,\sigma}+\rm{h.c} 
\,,\lb{eq:kin} \\
&&H_{\rm int}=U\sum_{m,i}n_{m,i,\uparrow}n_{m,i,\downarrow}\,, 
\ee
where $t$ (unit of energy) 
and $t'$ are the nearest-neighbor and inter-apex 
hoppings, respectively, $c_{m,i,\sigma}^\dagger$ creates 
a fermion with spin $\sigma$ 
on the $m$-th leg at the $i$-th unit cell, 
$n_{m,i,\sigma} = c_{m,i,\sigma}^\dagger c_{m,i,\sigma}$, 
and $U>0$ is the on-site repulsive interaction. 
Figure \ref{Fig1}(b) shows the band structure, 
$\epsilon(k)=\pm[4(1+\cos(k))+
(t')^2\cos^2(k)]^{1/2}+t'\cos(k)\,, 2t'\cos(k)$, 
in the noninteracting case ($U=0$). 
As we can see, 
one of the three bands becomes flat 
in the limit of $t'=0$ or $1$. 
We focus on the region where the filling 
of the whole bands is around $1/3$ 
(one fermion per unit cell on average) to 
investigate the effects of repulsive interaction.  
We have, for $t'= 0 \rightarrow 1$, a fully-occupied 
bottom band which touches the middle band at 
$k=\pm\pi$\,, where the middle (bottom) 
band becomes flat at $t'=0$ ($1$). 

Intriguingly, 
we have noticed in performing the DMRG 
that we have to keep an unusually large number of states up to $m_{\rm DMRG}=1500$ for the present ladder-like lattice. 
For DMRG we take an open boundary condition with inversion-symmetric configurations as shown in Fig.\ref{Fig1}(a)\,. 
Here we focus on the properties below $1/3$-filling to explore the possibility of fermion superfluidity 
in terms of the pair binding energy and correlation functions.

\begin{figure}[h]
\begin{center}
\includegraphics[width=1.0\linewidth]{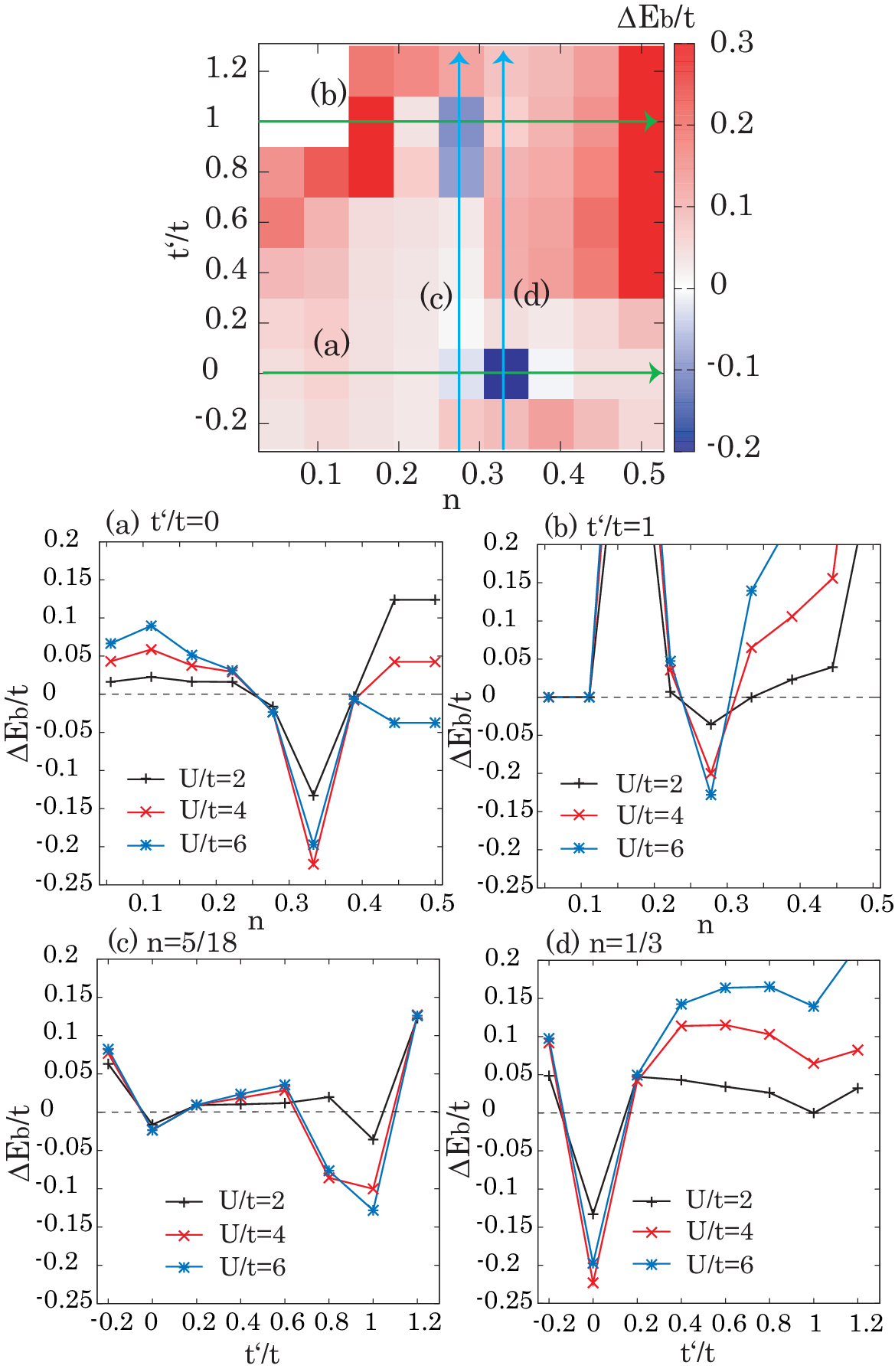}
\end{center}
\caption{(Color Online) 
(a,b) ED result for the binding energy $\Delta E_{\rm b}$ vs band filling $n$ 
for $t'=0$ (a) or $t'/t=1$ (b) for various values of $U/t$ 
with $N=18$ sites here.  
(c,d) Binding energy $\Delta E_{\rm b}$ vs $t'/t$ 
for band filling $n=5/18$ (c) or $n=1/3$ (d) for various values of $U/t$.  
Top panel is a color-code plot of $\Delta E_{\rm b}$ against 
$n$ and $t'/t$ for $U/t=4$, where arrows indicate the cross sections displayed 
in panels (a-d).
}
\label{Fig2}
\end{figure}

{\it Results---}  
Let us first examine the fermion pair formation in terms 
of the binding energy, 
$\Delta E_{b}\equiv E_{g}(N_\uparrow+1,N_\downarrow+1)+E_{g}(N_\uparrow,N_\downarrow) -2E_{g}(N_\uparrow+1,N_\downarrow)$, 
where $E_{g}(N_\uparrow,N_\downarrow)$ is the ground-state energy for 
$N_{\rm tot}=N_\uparrow+N_\downarrow$ fermions with $N_\sigma$ being 
the total number of $\sigma$-spin electrons. 
A negative $\Delta E_{b}$ implies that an attractive interaction works between two particles. 
$E_{g}(N_\uparrow,N_\downarrow)$ is computed with ED 
in periodic boundary conditions. 
In the numerical calculation, we set the total 
number of sites to be $N=18$ with 
the length of the chain being $L=N/3=6$\,.

Figure \ref{Fig2}(a) shows $\Delta E_{b}$ as a function of the filling $n = N_{\rm tot}/2N$ for  $t'=0$ for various values of $U/t$. 
We can immediately notice that two electrons 
become bound (i.e., $\Delta E_{b}$ becomes negative) 
sharply around $n = 1/3$ $(N_\uparrow=N_\downarrow=6\,, 
N=18)$
 for all the values of $U>0$ considered.  
Interestingly, the binding energy is not monotonic 
against $U$ but peaked around $U/t=4$.  As we shall see, 
the binding occurs for two electrons 
sitting on the $m=1$ and $3$-legs\,.
The binding energy continues to be negative just below $n=1/3$\,.
In the other flat-band limit at $t'/t=1$, 
we can see in Fig.\ref{Fig2}(b) 
that we have again a binding at 
a filling slightly smaller than 1/3 ($5/18$-filling), 
where $\Delta E_{b}$ becomes negative.

\begin{figure}[h]
\begin{center}
\includegraphics[width=1.0\linewidth]{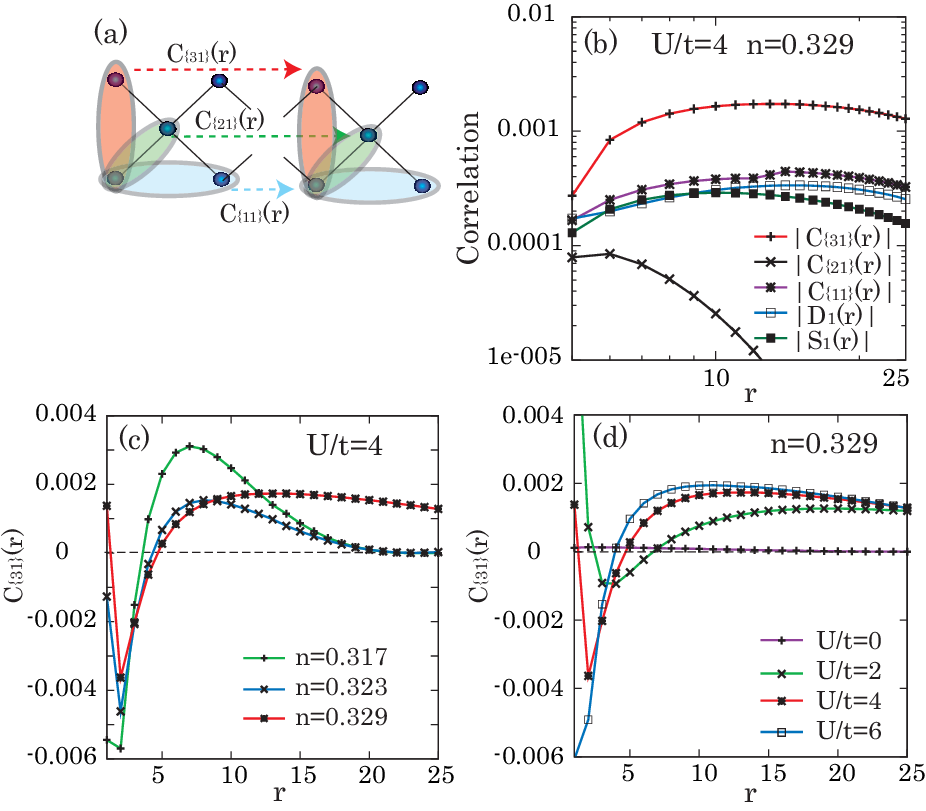}
\end{center}
\caption{(Color Online) (a) Correlation of various possible pair configurations on the diamond chain with $t'=0$. 
(b) Absolute values of various pair correlation functions are 
shown against real-space distance $r$ along with density and 
spin correlation functions for $U/t=4, n=0.329$\,. 
(c) Pair correlation $C^{\rm pair}_{\{31\}}(r)$ against $r$ 
for various values of $n$ for  $U/t=4$.  
(d) Pair correlation $C^{\rm pair}_{\{31\}}(r)$ for various values of 
$U/t$ for $n=0.329$.  
Here the length of the chain is $L=55$ (with 164 sites in total).
}
\label{Fig3}
\end{figure}

We now proceed to DMRG calculations for 
various correlation functions, including pair correlation, 
on the diamond chain \cite{ave}. 
The density ($D_{m}$) and spin ($S_{m}$) correlation functions on 
the $m$-th leg are defined respectively as
\be
&&D_{m}(i,j)=\langle n_{m,i}n_{m,j}\rangle-\langle n_{m,i}\rangle\langle n_{m,i}\rangle\,, \\
&&S_{m}(i,j)=\langle S_{m,i}^{(z)}S_{m,j}^{(z)}\rangle\,,
\ee
with $n_{m,i}= n_{m,i,\uparrow}+n_{m,i,\downarrow}$ and 
$S_{m,i}^{(z)}=(n_{m,i,\uparrow}-n_{m,i,\downarrow})/2$\,. 
We compute the correlation functions on leg $m=1$ and on $2$ 
(while the correlation functions on $m=3$ is equivalent to those on $m=1$). 
The singlet-pair correlation functions are defined as 
\be
&&C_{\{m'm\}}^{\rm pair}(i,j)=\langle\Delta_{m'm,j}\Delta_{m'm,i}^{\dagger}\rangle\,, \\
&&\Delta_{m'm,i} \equiv c_{m',i+l,\uparrow}c_{m,i,\downarrow}-c_{m',i+l,\downarrow}c_{m,i,\uparrow}\,, 
\ee
where $l$ characterizes the pair [see Fig.\ref{Fig3}(a)].

The result for various correlations in Fig.\ref{Fig3}(b) reveals that 
the dominant (most long-tailed with distance $r$) 
correlation for $U/t\neq 0$ in the vicinity of $1/3$-filling 
($n\simeq 0.329$ with 
$N_{\uparrow}=N_{\uparrow}=54$ and 
$N=164$)
is the pair correlation $|C^{\rm pair}_{\{31\}}(r)|$ for the pair, 
\[
\Delta_{31,i}= c_{3,i,\uparrow}c_{1,i,\downarrow}-c_{3,i,\downarrow}c_{1,i,\uparrow},
\] 
across $m=1$ and $3$ (see Fig.\ref{Fig3}(a))\,.\cite{paircomment} 
The next dominant correlations 
are the pair $C^{\rm pair}_{\{11\}}(r)$ 
(for $\Delta_{11,i}=c_{1,i+1,\uparrow}c_{1,i,\downarrow}-c_{1,i+1,\downarrow}c_{1,i,\uparrow}$) and 
density $D_{1}(r)$ correlations on $m=1$\,.  
Then comes the spin $S_{1}(r)$ correlation on $m=1$. 
On the other hand, the correlations on $m=2$ 
(see $C^{\rm pair}_{\{21\}}(r)$ and Supplementary Material B) rapidly decay for all the values of $n$ studied here \cite{ave}. 
As in the density and spin correlations, the pair 
correlation involving $m=2$ 
($\Delta_{21,i}=c_{2,i,\uparrow}c_{1,i,\downarrow}-c_{2,i,\downarrow}c_{1,i,\uparrow}$) 
shows a fast decay. 
The dominant $\Delta_{31,i}$ is consistent with
an analysis of the entanglement entropy and edge states at $t'=0$\, in Supplementary Material A.

The reason why all of the pair, density and spin correlations  develop on legs $m=1,3$ in the vicinity of $1/3$-filling 
can be considered as coming from the basis 
functions on the flat band. 
When the hopping $t'$ is absent, we can introduce 
a basis, 
\be
&&\alpha_{i,\sigma}=c_{2,i,\sigma} \,, \quad
\beta_{i,\sigma}=(c_{1,i,\sigma}+c_{3,i,\sigma})/\sqrt{2}\,, \nn\\
&&\gamma_{i,\sigma}=(c_{1,i,\sigma}-c_{3,i,\sigma})/\sqrt{2} \,,\lb{eq:trans}
\ee
with which the kinetic part of Eq.(\ref{eq:hami}) can be 
expressed as
$H_{{\rm kin},t'=0}=\sqrt{2}t\sum_{i,\sigma}
\alpha_{i,\sigma}^\dagger (
\beta_{i,\sigma}+\beta_{i-1,\sigma})
+\rm{h.c} $\,. 
The basis $\{\gamma_{i,\sigma}\}$ represents the particles on 
the flat band (see left panel of Fig.\ref{Fig1}(c)), 
in which the probability amplitude selectively resides on legs $m=1$ and $3$ (i.e. on A sublattice if we divide the 
bipartite lattice). 
The interaction $U$ then brings about interband matrix 
elements between the flat and dispersive bands around $1/3$-filling. 
The development of superconductivity when 
the flat band is empty (which is equivalent 
to full filling in the present electron-hole symmetric 
lattice) is consistent with the result in 
Ref.\cite{Kuroki}.   
While the latter uses FLEX, a 
weak-coupling method, the present result reveals 
the flat-band superconductivity is in fact prominent 
in a strong-coupling ($U/t\simeq 4$) regime. 
The behavior of the correlation functions 
enhanced on $m=1,3$ should come from 
the virtual states that have probability amplitudes 
residing on legs $m=1$ and $3$ with 
the long-range nature of the correlations 
involving orbits for the flat band.

What happens when the filling 
is exactly $1/3$ is also interesting, 
so that we have studied the quantum phases 
at that filling 
in Supplemental Material A.  Topological states 
are shown to emerge, which 
is indicated from the entanglement spectrum 
for spins on the outer sites as well 
as from emerging edge states. This is considered 
to be another 
effect of the unusual Wannier states in the 
flat band, and the pairing states for the 
$E_F$ close to but away from the 
flat band seems to sits adjacent to 
a topological phase at the point where 
the flat band just becomes empty.

{\it Summary---} 
We have investigated repulsively interacting fermions 
on the diamond chain, a simplest possible quasi-1D 
flat-band system, with ED and DMRG calculations. 
The numerical results have revealed that when 
the band filling is slightly below $1/3$ with 
the flat band close to but away from $E_F$, 
the pair binding energy calculated with ED 
has two sharp peaks 
at two flat-band limits ($t'=0$ or $1$). 
Then the DMRG shows that, for $t'=0$, 
the most dominant correlation is 
the singlet-pair across the outer sites 
($m=1,3$) of the diamond. 
For $t'/t=1$, by constrast, a phase separated behavior 
is observed 
as indicated in Supplementary Material B.
The flat band promoting superconductivity 
through virtual pair hoppings involving the 
band as conceived in FLEX\cite{Kuroki} 
is shown to be prominent in a strong-coupling regime.  
It is an interesting future problem to see whether 
a mechanism beyond the boson-exchange is at 
work here, which will require methods that 
take account of vertex corrections.  

While we have concentrated on the quasi-1D diamond chain, 
enhanced pairing correlations with 
the major component residing on the flat-band 
wave functions are expected to be a general property 
of the flat-band systems satisfying the connectivity 
condition.  
Extension of the present study 
to flat-band systems with fluxes 
inserted\cite{Vidal,vollhardt,Takayoshi} 
is also an interesting future work. 
While the diamond-chain structure has been 
discussed for condensed-matter systems such 
as an insulating magnet azurite\cite{azurite1,azurite2},  
cold atoms on optical lattices should be an ideal 
test bench for experimental realizations of 
flat-band lattices. 

\begin{acknowledgments}
HA was supported by a JSPS KAKENHI (No. 26247057) 
and ImPACT project (No. 2015-PM12-05-01) from JST.  
SY and MM were partially supported by JSPS KAKENHI (Nos.
15K00178, 26400322, and 16H02450).  
The numerical work was in part performed on Fujitsu BX900 at JAEA. 
\end{acknowledgments}

\section*{Supplementary Material}

\section*{A. Insulating phases and edge states at $1/3$-filling}

\begin{figure}[h]
\begin{center}
\includegraphics[width=1.0\linewidth]{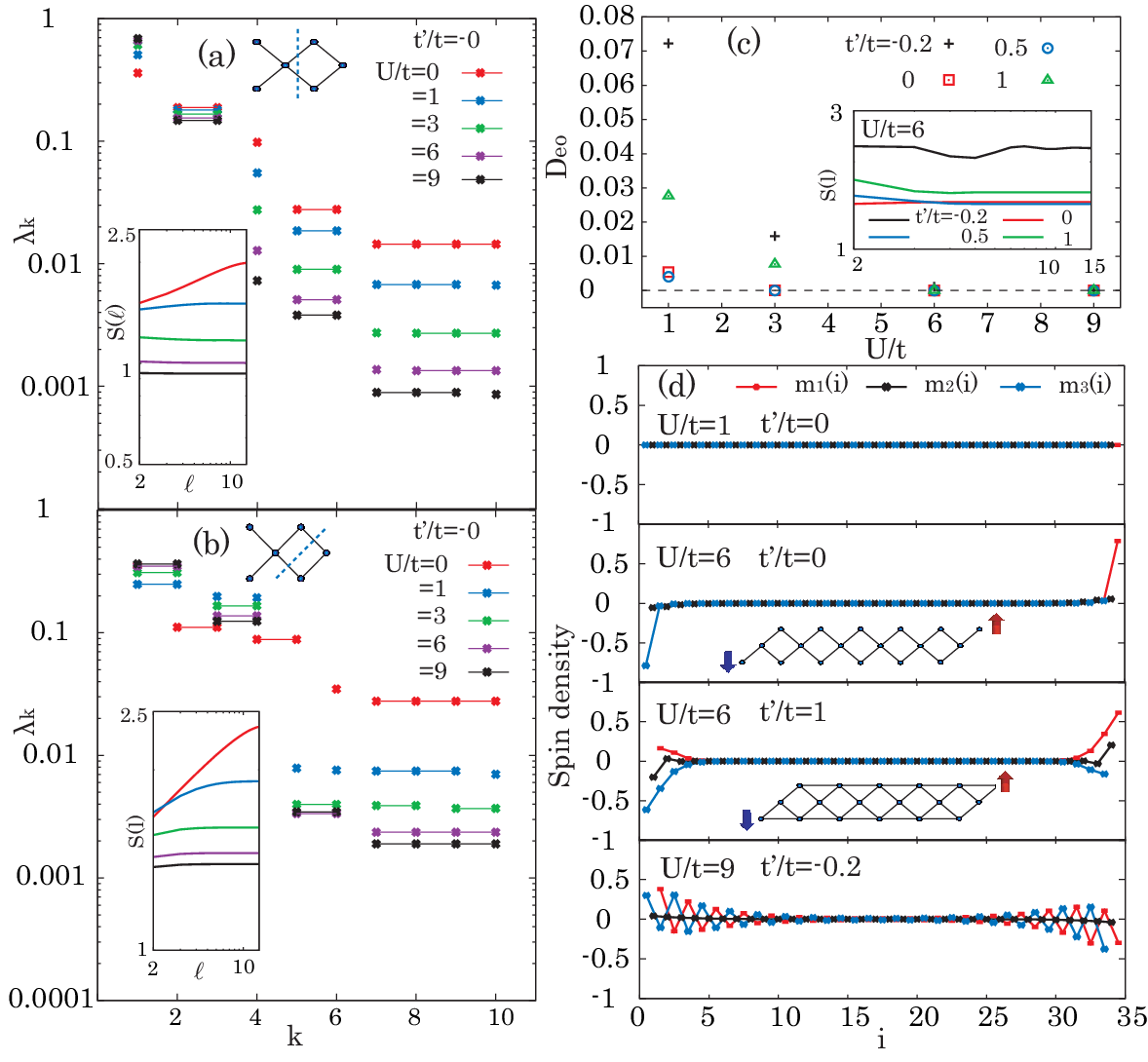}
\end{center}
\caption{(Color Online) 
(a) and (b) display eigenvalues of the 
reduced density matrix $\lambda_{k}$ versus the eigenvalue 
number $k$ 
for vertical(a) and diagonal(b) cuts for various values of $U/t$ with  $t'=0$. 
Degenerate values are connected by horizontal lines.  
Respective insets show the entanglement entropy $S(\ell)$ 
against the block length $\ell$ on double logarithmic scales 
for $t'=0$.  
(c) $D_{\rm eo}=\sum_{k=1}^{N=100}(\lambda_{2k}-\lambda_{2k-1})/\lambda_{2k-1}$ (see text) 
against the interaction $U/t$ for the diagonal cut 
for various values of $t'$.  
Inset shows the corresponding entanglement entropy 
$S(\ell)$ against $\ell$ for $U/t=6$\,.
(d) Spin density profiles for, from top to bottom, 
$U/t=1, t'=0$; $U/t=6, t'=0$; $U/t=6, t'/t=1$ and $U/t=9, t'/t=-0.2$ for diagonal cuts.
}
\label{FigS1}
\end{figure}

In this section, 
we present the results for the diamond chain exactly at $1/3$-filling by evoking 
entanglement entropy and spectrum analysis. 
Entanglement entropy is a useful tool for distinguishing 
critical from gapped phases\cite{Vidal2003,Korepin2004}. 
If a finite block of length $\ell$ is considered on 
a chain of $L$ sites, 
the reduced density matrix is defined as $\rho_{L}(\ell)={\rm Tr}_{L-\ell} |\Psi\rangle\langle\Psi|$, 
and the corresponding entanglement entropy (von Neumann entropy) is given by
\be
S_{L}(\ell)=-{\rm Tr}\rho_{L}(\ell)\ln\rho_{L}(\ell) \,.
\ee
As well-known\cite{Vidal2003,Korepin2004}, if the system 
has a gapless excitation spectrum, the 
entanglement entropy should grow logarithmically with the 
block size $\ell$\,. 
On the other hand, when all the excitations are gapped, $S_{L}(\ell)$ should tend to a constant for large $\ell$\,. 
Furthermore, the eigenvalues (entanglement spectrum; ES) of the equation for the reduced density matrix, 
$\rho_{L}\bo{u}_{k}=\lambda_{k}\bo{u}_{k}$, 
contain rich 
information on bulk properties. 
For instance, degeneracies of ES determine the parity of the many-body wavefunctions and
distinguishes nontrivial topological states from trivial product states\cite{Pollmann2010,Pollmann2012}. 
In addition, doubly degenerate ES in the bulk indicates doubly degenerate edge spectrum in finite systems.

First, we focus on the original diamond chain with 
$t'=0$ for 
filling $n=\frac{N_{\rm tot}}{2N}\simeq 0.337$, 
where $N_{\rm tot} (=70$ here) is the total number of 
particles and $N (=104)$ is the total number of sites. 
The filling  $0.337$ corresponds to $1/3$-filling in 
the periodic boundary condition, 
where the bottom dispersive band is fully occupied and 
touches the flat band at $k=\pm \pi$ (see Fig.1(b) in the main text)\,.  
In calculating the entanglement entropy, we can introduce two types of cuts of the diamond chain, 
namely vertical and diagonal cuts as shown in Fig.1(a) in the main text.  
Insets of Fig.\ref{FigS1}(a,b) show the entanglement entropy $S(\ell)$ 
against the block length $\ell$ for the vertical and diagonal cuts. 
We can see in both cases that the initial slope of $S_{L}(\ell)$ vs 
$\ell$ for $U/t=0$ 
shows a logarithmic growth as a consequence of gapless excitations. 
As $U/t$ is increased, $S_{L}(\ell)$ rapidly saturates to a constant, 
which implies a phase transition from gapless to gapped phases induced by repulsive $U$\,. 
Note that this phase transition is distinct from the Mott transition at half-filling: 
All the excitation spectra are gapped and transition occurs at $1/3$-filling. 

If we turn to eigenvalues of the reduced density matrix 
(entanglement spectrum), 
it is known that ES can depend on how the 
system is divided into 
subsystems\cite{Hida2016,Arikawa2009}.  
Figures \ref{FigS1}(a) and (b) respectively 
show the ES for the vertical and diagonal cuts, 
where we incise the system at the center of the chain. 
We find that the ES for the vertical cut has no even degeneracies for 
all the values of the interaction studied here, 
whereas even degeneracies appear for the diagonal cut 
when the interaction $U/t$ becomes large. 
ES behavior goes hand in hand 
with edge states, which is known as the bulk-edge 
correspondence. 
The spin density profile in Fig.\ref{FigS1}(d) reveals that, while ES degeneracies are 
not developed for $U/t=1$ (see the values at $k=9,10$ in Fig.\ref{FigS1}(a)), 
edge states (accommodating free $1/2$-spins) do emerge around 
the diagonally-cut boundaries 
for a larger $U/t=6$, where the even degeneracies are 
clearly seen. In this sense the diagonal cut here 
may be similar to cutting a spin singlet in 
the Haldane model\cite{AKLT}.

Next, we turn to the case with $t'\neq 0$\,. 
Let us define a quantity $D_{\rm eo}=\sum_{k=1}^{N}(\lambda_{2k}-\lambda_{2k-1})/\lambda_{2k-1}$ as a measure 
of the even-odd degeneracy of ES: 
$D_{\rm eo}$ should vanish for even degeneracies. 
In Fig.\ref{FigS1}(c), the even degeneracy for diagonal cut can be observed for all the values of $t'$ 
for strong enough interaction $U/t$\,. 
As in the $t'=0$ case, the degeneracy is seen to arise when 
the gapped phase is formed, 
where the entanglement entropy $S(\ell)$ becomes 
constant for large $\ell$ [see inset of Fig.\ref{FigS1}(c)]\,. 
We note that even degeneracies for vertical cut do not emerge as in 
$t'=0$. In lower panels in Fig.\ref{FigS1}(d) 
the edge states derived from the even degeneracies 
are seen to exhibit various structures depending on the value of 
$t'$\,: 
Edge states for $t'=1$ broaden over several lattice 
sites and 
across different legs ($m$), in contrast to the 
$t'=0$ case where edge states are highly localized at $m=1$ or $m=3$ around the boundary. 
By contrast, when $t'= -0.2 <0$, for which we have 
Fermi points (see Fig.1(b) in the main text), 
edge states on $m=1$ or $3$ have a staggered magnetization with an exponential 
decay off the edge.

Thus we find that the edge states take various structures depending on the value of $t'$\,. 
In all cases, even degeneracies appear only for diagonal cut. 
This means that the entanglement pair is mainly formed 
across the legs $1$ and $3$.  
Specifically, the edge states for $t'/t=0$ are sharply localized at the diagonal edge boundary, which 
suggests that the entanglement pair for $t'=0$ form 
dimer-like states across legs $1$ and $3$.

\section*{B. Density profiles and correlations
below $1/3$-filling}

\begin{figure}[h]
\begin{center}
\includegraphics[width=1.0\linewidth]{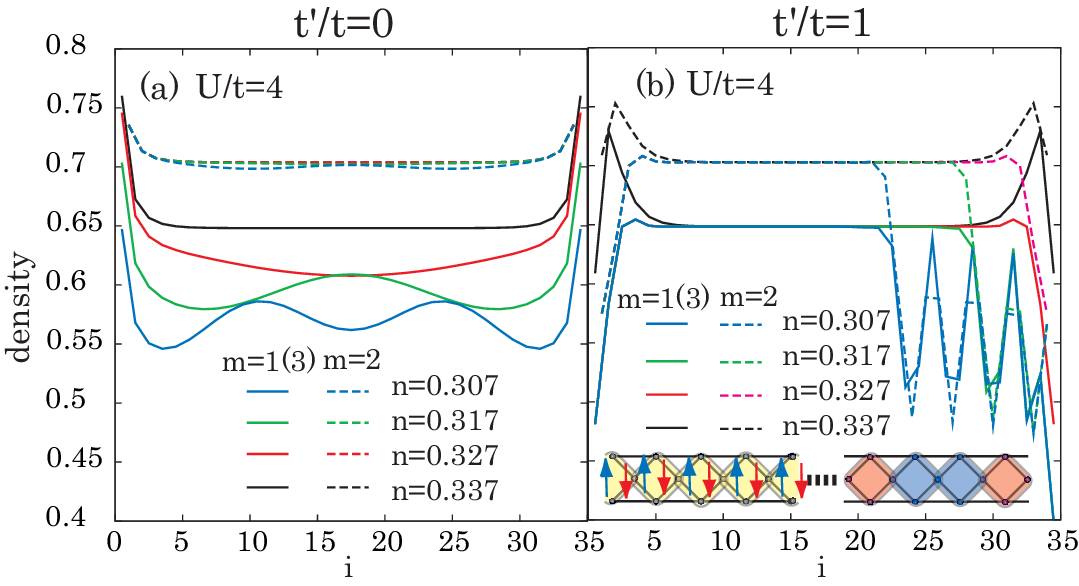}
\end{center}
\caption{(Color Online) Density profiles 
on legs $m=1,3$ (solid lines) or on $m=2$ (dashed) 
for various values of $n\leq 1/3$ with $U/t=4$ 
for $t'=0$ (a) or $t'/t=1$ (b). 
The system has open boundaries with length $L=35$ (with $N$=104 sites in total). 
The lower band is fully occupied at filling $0.337$, which corresponds to $1/3$-filling in the periodic boundary.
Insets in (b) schematically represent phase separation between plateau region (left inset) and CDW-like (right) region. 
The plateau is formed by doubly-occupied states in 
the flat-band basis, 
which corresponds to those in the insulating phase at $1/3$-filling, 
while the CDW-like to a configuration of the localized states with $2\pi/3$-periodicity. 
Inset of (b) systematically show the CDW-like configuration, where high (red) and low (blue) density regions 
are aligned with $2\pi/3$-periodicity.
}
\label{FigS2}
\end{figure}

\begin{figure*}[h]
\begin{center}
\includegraphics[width=1.0\linewidth]{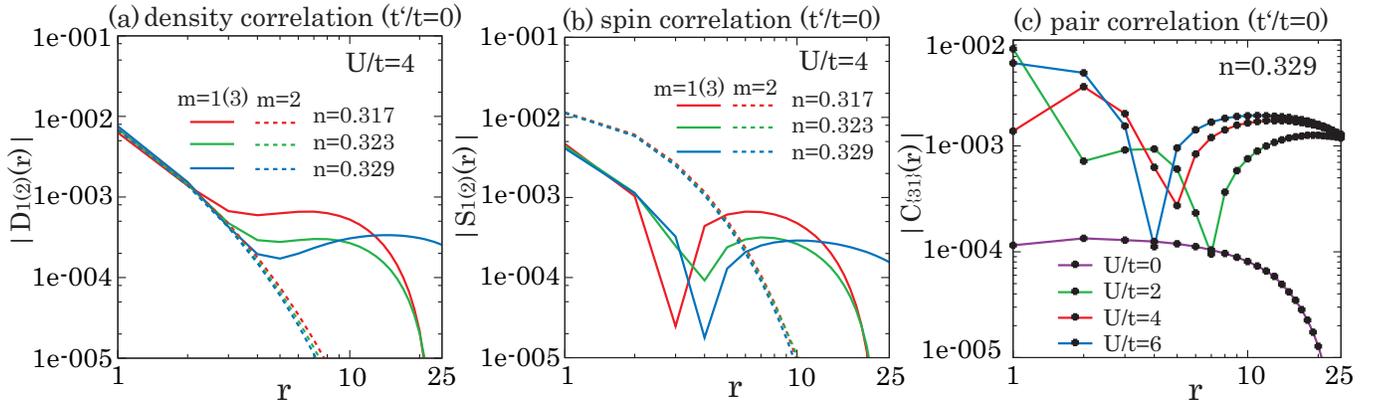}
\end{center}
\caption{(Color Online) The absolute value of 
the density (a) and spin (b) correlation functions 
on legs $m=1,3$ (solid lines) or on $m=2$ (dashed)  
for various values of $n<1/3$ with $t'/t=0$ and $U/t=4$. 
(c) is the absolute value of the pair correlation functions $C_{31}(r)$ for various values of $U/t$ with $t'/t=0$ and $n=0.329$. 
All figures are shown on log-log scales.
Length of the chain is $L=55$ with $N=165$ sites in total 
and $i_0= L/4 $.
}
\label{FigS3}
\end{figure*}

In order to identify a phase separation (as another pair-binding 
effect distinct from superfluidity), 
let us show in Fig.\ref{FigS2} the density profiles calculated with DMRG for open boundary conditions.  
For $t'=0$ density profiles below $1/3$-filling only show Friedel-like oscillations induced by boundaries.
For $t'/t=1$, by contrast, phase separation between 
plateau and CDW-like regions is observed.  
We can analyze this by first noting 
that particles on the flat band for $t'/t=1$ can be represented by 
a basis comprising localized states 
(see right inset of Fig.1 in the main text),
\be
b_{\sigma,i}=c_{2,\sigma,i}+c_{2,\sigma,i+1}-c_{1,\sigma,i+1}-c_{3,\sigma,i+1}\,,
\ee 
with which we have 
\be
&&|\Psi\rangle=\prod_{\sigma,i}(b_{\sigma,i}^\dagger)^{n_{\sigma,i}}|0\rangle \,, \\
&&H_{{\rm kin},t'=1}|\Psi\rangle=-2t\sum_{\sigma,i}n_{\sigma,i} |\Psi\rangle\,, 
\ee
where $n_{\sigma,i}$ is the particle number in the 
localized state $i$\,. 
The CDW-like region corresponds to a configuration of the localized states $b_{\sigma,i}$ with $2\pi/3$-periodicity, while 
the plateau region represents doubly-occupied states, 
$b_{\downarrow,i}^\dagger b_{\uparrow,i}^\dagger|0\rangle$, 
which is the gapped phase formed at $1/3$-filling. 
Although the result for the binding energy (Fig.1 in the main text) shows 
a two-particle attraction for $t'/t=1$, 
the phase separation observed here implies 
formation of domains. 
This contrasts with the case of $t'=0$, where we have 
an attractive binding energy but with 
no phase-separated behavior, which 
is consistent with formation of itinerant pairs 
as seen in the pair correlation. 

We also display in Fig.\ref{FigS3} the density and spin correlation functions as well as pair
correlation, on double-logarithmic scales for various values of $n$ here.  As before, correlation 
functions are calculated for the interior of the finite system. 
While we have some dip structures due to spatial modulations, the overall decay of the density $D_{1}(r)$ and spin $S_{1}(r)$
correlations become slow in the vicinity of $1/3$-filling, but not so slowly 
decaying as the pair correlation.   We can again confirm that  the behavior is quite sensitive to the values of $n$ and  $U/t$

\section*{C. Hamiltonian in the ``Wannier" basis}
It is curious how the Hamiltonian, Eq.(1) in the main text, should look like 
when expressed in ``Wannier" basis. 
For the original diamond chain with $t'=0$, we can introduce the basis 
$
\alpha_{i,\sigma}=c_{2,i,\sigma}
$, 
$\beta_{i,\sigma}=(c_{1,i,\sigma}+c_{3,i,\sigma})/\sqrt{2}
$  for the dispersive bands, and 
$
\gamma_{i,\sigma}=(c_{1,i,\sigma}-c_{3,i,\sigma})/\sqrt{2}
$\, for the flat band.   
The interaction term $H_{{\rm int}}$ can then be expressed as
\be
H_{{\rm int}}&=&\sum_{i}h_{{\rm int},i}\,, \\
h_{{\rm int},i}&=&\frac{U}{2}
\sum_{\lambda,\lambda'=\beta,\gamma}n_{\lambda,\uparrow,i}n_{\lambda',\downarrow,i}
+Un_{\alpha,\uparrow,i}n_{\alpha,\downarrow,i} \nn \\
&&+\frac{U}{2}\left(\rho_{\beta,i}^{(+)}\rho_{\gamma,i}^{(-)}-S_{\beta,i}^{(+)}S_{\gamma,i}^{(-)}+\rm{h.c.}\right) 
\,, 
\ee
where, for $\lambda=\beta,\gamma$, $n_{\lambda,\sigma,i}=\lambda_{\sigma,i}^\dagger \lambda_{\sigma,i}$, while 
$S_{\lambda,i}^{(-)}=\lambda_{\downarrow,i}^\dagger \lambda_{\uparrow,i}$ and $S_{\lambda,i}^{(+)}=[S_{\lambda,i}^{(-)}]^\dagger$
are spin-flip operators, and 
$\rho_{\lambda,i}^{(-)}=\lambda_{\uparrow,i}\lambda_{\downarrow,i}$ and $\rho_{\lambda,i}^{(+)}=[\rho_{\lambda,i}^{(-)}]^\dagger$
pair-hopping operators.  
The pair-hopping interaction, $\rho_{\beta,i}^{(+)}\rho_{\gamma,i}^{(-)}$, 
allows a doubly occupied state, 
$
|\Psi\rangle_{i}=
(\beta_{\uparrow,i}\beta_{\downarrow,i}-
\gamma_{\uparrow,i}\gamma_{\downarrow,i}
)^\dagger|0\rangle_{i}
$, 
to be the lowest eigenstate of an isolated unit cell Hamiltonian, 
$h_{{\rm int},i}|\Psi\rangle_{i}=0$, at $1/3$-filling. 
This is expected to favor pair formation. 
Indeed, $
\beta_{\uparrow,i}\beta_{\downarrow,i}-
\gamma_{\uparrow,i}\gamma_{\downarrow,i}
$ is nothing but the singlet pair in the original bases, 
$
\Delta_{31,i}= c_{3,i,\uparrow}c_{1,i,\downarrow}-c_{3,i,\downarrow}c_{1,i,\uparrow}
$ considered in the main text.

\section*{D. Binding energy for various values of $U/t$}
\begin{figure*}[h]
\begin{center}
\includegraphics[width=1\linewidth]{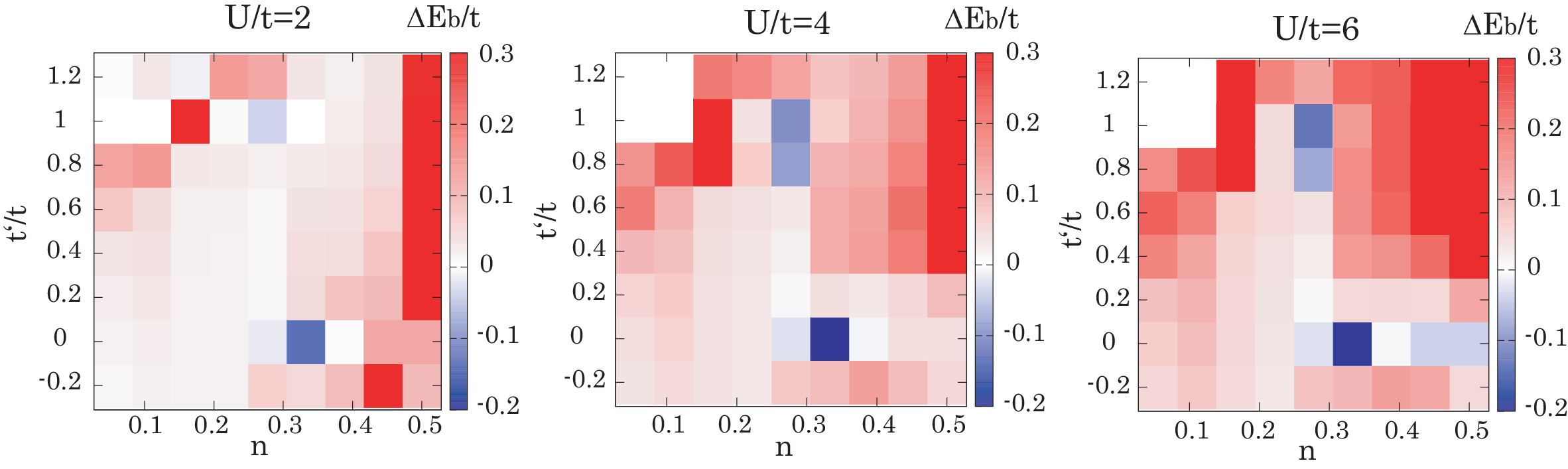}
\end{center}
\caption{(Color Online) Color-code plot of $\Delta E_{\rm b}$ against 
$n$ and $t'/t$ for various values of $U/t$.
}
\label{FigS4}
\end{figure*}

Figure \ref{FigS4} plots the binding energy against the density 
$n$ and $t'/t$ for various values of $U/t$.  
Attractive binding energies are found also above $1/3$-filling at $U/t=6$, 
but development of pair correlation is not found in ED and DMRG calculations 
for the sample sizes treated here. 

\begin{thebibliography}{99}


\bibitem{Lieb}
E.~H.~Lieb, Phys. Rev. Lett. {\bf 62}, 1201 (1981).
\bibitem{Mielke}
A.~Mielke, J. Phys, A: Math. Gen. {\bf 24}, L73 (1991); {\it ibid} {\bf 24}, 3311 (1991).
\bibitem{Tasaki}
H.~Tasaki, Phys. Rev. Lett. {\bf 69}, 1608 (1992).

\bibitem{tasakiPTP}
H. Tasaki, 
Prog. Theore. Phys. {\bf 99}, 489 (1998). 

\bibitem{yamada} See, e.g., refs in M. Yamada, T. Soejima, N. Tsuji, D. Hirai, M Dinca and H. Aoki, Phys. Rev. B {\bf 94}, 081102(R) (2016), which proposes a design of a flat-band 
ferromagnetism in an organometallic system.

\bibitem{TakahashiLieb} 
S. Taie, H. Ozawa, T. Ichinose, T. Nishio, S. Nakajima and Y. Takahashi, 
Science Advances {\bf 1}, e1500854 (2015).

\bibitem{Kagome}
G.-B.ong Jo, et al. Phys. Rev. Lett. {\bf 108}, 045305 (2012).


\bibitem{Tang}
E.~Tang, J.-W.~Mei, and X.-G.~Wen, 
Phys. Rev. Lett. {\bf 106}, 236802 (2011).
\bibitem{Sun}
K.~Sun, Z.~Gu, H.~Katsura, and S.~Das Sarma, 
Phys. Rev. Lett. {\bf 106}, 236803 (2011).
\bibitem{Neupert}
T.~Neupert, L.~Santos, C.~Chamon, and C.~Mudry, 
Phys. Rev. Lett. {\bf 106}, 236804 (2011).
\bibitem{Li}
X.~Li, E.~Zhao, and W.~Vincent Liu, 
Nat. Commun. {\bf 4}, 1523 (2013).

\bibitem{Tovmasyan} M. Tovmasyan et al, arXiv:1608.00976 
study the attractive Hubbard model to 
discuss the superfluuid weight of a flat band. 

\bibitem{RVB} 
R.~R.~Montenegro-Filho and M.~D.~Coutinho-Filho, 
Phys. Rev. B {\bf 74}, 125117 (2006).

\bibitem{Vidal}
J.~Vidal, B.~Dou\c{c}ot, R.~Mosseri, and P.~Butaud,  
Phys. Rev. Lett. {\bf 85}, 3906 (2000); 
B.~Dou\c{c}ot and J.~Vidal, {\it ibid} {\bf 88}, 227005 (2002). 

\bibitem{Kuroki} K. Kuroki, T. Higashida and R. Arita, 
Phys. Rev. B {\bf 72}, 212509 (2005).

\bibitem{Takayoshi}
S. Takayoshi, H. Katsura, N. Watanabe, and H. Aoki, 
Phys. Rev. A {\bf 88}, 063613 (2013). 
\bibitem{Huber}
M. Tovmasyan, E. P. L. van Nieuwenburg, and S. D. Huber,
Phys. Rev. B {\bf 88}, 220510(R) (2013). 


\bibitem{DMRG1}
S.~R.~White, Phys.~Rev.~Lett. {\bf 69}, 2863 (1992);
 Phys.~Rev.~B {\bf 48}, 10345 (1993).
\bibitem{DMRG2}
K.~A.~Hallberg, Adv.~Phys. {\bf 55}, 477 (2006); 
U. Schollw\"{o}ck, Ann.~Phys. {\bf 326}, 96 (2011).
\bibitem{Okumura2009}
M.~Okumura, S.~Yamada, M.~Machida, and T.~Sakai, 
Phys. Rev. A {\bf 79}, 061602(R) (2009).
\bibitem{Okumura2011}
M.~Okumura, S.~Yamada, M.~Machida, and H.~Aoki, 
Phys. Rev. A {\bf 83}, 031606(R) (2011).

\bibitem{paircomment} The magnitude of the pair 
correlation $\sim 10^{-3}$ is similar to those 
numerically calculated for the standard ladder systems such as the three-leg Hubbard ladder, see, e.g., 

T. Kimura, K. Kuroki and H. Aoki, 
Phys. Rev. B {\bf 54}, R9608 (1996); 
J. Phys. Soc. Jpn {\bf 66}, 1599 (1997); 
{\it ibid} {\bf 67}, 1377 (1998).

\bibitem{ave}
To minimize the effect of edges, 
we calculate correlation functions for 
an interior region as,\cite{Alejandro} 
$C(r)=\frac{1}{L-2i_0-r}\sum_{i_0<i}^{L-i_0-r}C(i+r,i)$, 
where $C(i+r,i)$ stands for the two-point correlation function 
and $i_0 (=L/4$ in Fig.\ref{Fig3}) is set well away from the sample edges.
\bibitem{Alejandro}
A. M.~Lobos, M. Tezuka, and A. M. Garcia-Garcia, 
Phys. Rev. B {\bf 88}, 134506 (2013).

\bibitem{vollhardt} 
Z.~Gulacsi, A.~Kampf,and D.~Vollhardt, 
Phys. Rev. Lett. {\bf 99}, 026404 (2007).


\bibitem{azurite1} 
H.~Kikuchi, Y.~Fujii, M.~Chiba, S.~Mitsudo, T.~Idehara, 
T.~Tonegawa, K.~Okamoto, T.~Sakai, T.~Kuwai, and H.~Ohta,  
Phys. Rev. Lett. {\bf 94}, 227201 (2005).
\bibitem{azurite2}
H. Jeschke, I. Opahle, H. Kandpal, R. Valenti, H. Das, T. Saha-Dasgupta, O. Janson, H. Rosner, A. Br\"uhl, B. Wolf, M. Lang, J. Richter, S. Hu, X. Wang, R. Peters, T. Pruschke, and A. Honecker, Phys. Rev. Lett. {\bf 106}, 217201 (2011).










\end{thebibliography}

\begin{thebibliography}{99}
\bibitem{Vidal2003}
G.~Vidal, J.~I.~Latorre, E.~Rico, and A.~Kitaev, 
Phys. Rev. Lett. {\bf 90}, 227902 (2003). 
\bibitem{Korepin2004}
V.~E.~Korepin, 
Phys. Rev. Lett. {\bf 92}, 096402 (2004).
\bibitem{Pollmann2010}
F.~Pollmann, A.~M.~Turner, E.~Berg, and M.~Oshikawa, 
Phys. Rev. B {\bf 81}, 064439 (2010).
\bibitem{Pollmann2012}
F.~Pollmann, E.~Berg, A.~M.~Turner, and M.~Oshikawa, 
Phys. Rev. B {\bf 85}, 075125 (2012). 
\bibitem{AKLT}
I.~Affleck, T.~Kennedy, E.~H.~Lieb, and H.~Tasaki,
 Phys. Rev. Lett. {\bf 59}, 799 (1987);
 Commun. Math. Phys. {\bf 115}, 477 (1988).
 
\bibitem{Hida2016}
K. Hida, 
J.~Phys.~Soc.~Jpn.~{\bf 85}, 024705 (2016).
\bibitem{Arikawa2009}
M. Arikawa, S. Tanaka, I. Maruyama, and Y. Hatsugai, 
Phys. Rev. B {\bf 79}, 205107 (2009). 
\end{thebibliography}
\end{document}